\documentstyle[eqsecnum,multicol,aps]{revtex}

\begin{document}

\draft 
\input epsf

\title{Excitons in T-shaped quantum wires}     

\author{M. H. Szymanska, P. B. Littlewood \thanks{and Bell
Laboratories, Lucent Technologies, Murray Hill, NJ 07974 USA} and
R. J. Needs }

\address{ Theory of Condensed Matter, Cavendish Laboratory, Cambridge
CB3 0HE, UK }  

\date{\today} 

\maketitle

\begin{abstract}
We calculate energies, oscillator strengths for radiative
recombination, and two-particle wave functions for the ground state
exciton and around 100 excited states in a T-shaped quantum wire. We
include the single-particle potential and the Coulomb interaction
between the electron and hole on an equal footing, and perform exact
diagonalisation of the two-particle problem within a finite basis
set. We calculate spectra for all of the experimentally studied cases
of T-shaped wires including symmetric and asymmetric
GaAs/Al$_{x}$Ga$_{1-x}$As and
In$_{y}$Ga$_{1-y}$As/Al$_{x}$Ga$_{1-x}$As structures. We study in
detail the shape of the wave functions to gain insight into the nature
of the various states for selected symmetric and asymmetric wires in
which laser emission has been experimentally observed. We also
calculate the binding energy of the ground state exciton and the
confinement energy of the 1D quantum-wire-exciton state with respect
to the 2D quantum-well exciton for a wide range of structures, varying
the well width and the Al molar fraction $x$. We find that the largest
binding energy of any wire constructed to date is 16.5 meV. We also
notice that in asymmetric structures, the confinement energy is
enhanced with respect to the symmetric forms with comparable
parameters but the binding energy of the exciton is then lower than in
the symmetric structures.  For GaAs/Al$_{x}$Ga$_{1-x}$As wires we
obtain an upper limit for the binding energy of around 25 meV in a 10
{\AA} wide GaAs/AlAs structure which suggests that other materials
must be explored in order to achieve room temperature applications.
There are some indications that
In$_{y}$Ga$_{1-y}$As/Al$_{x}$Ga$_{1-x}$As might be a good candidate.
\end{abstract}

\begin{multicols}{2}
\section {Introduction} 
\label{Intr}

Optical properties of electrons and holes confined to few dimensions
are of interest for optical and electronic devices.  As the
dimensionality of the structure is reduced, the density of states
tends to bunch together leading to a singularity in the 1D case. This
effect can be very useful for low-threshold laser applications.  At
the same time the excitonic interaction in 1D is enhanced with respect
to that in 3D and 2D structures. Quantum confinement leads to an
increase in the exciton binding energy, $E_b$, and the oscillator
strength for radiative recombination.  Both effects provide
possibilities for much better performance of optical devices such as
semiconductor lasers.
                       
The binding energy of a ground-state exciton in an ideal 2D quantum
well is four times that in the 3D bulk semiconductor.  For the ideal
1D quantum wire $E_b$ diverges. This suggests that $E_b$ for quasi-1D
wires can be greatly increased with respect to the 2D limit for very
thin wires with high potential barriers.  3D and 2D excitons
dissociate at room temperature to form an electron-hole plasma. To
make them useful for real device applications, their binding energy
needs to be increased and this might be achieved by using 1D quantum
confinement.

Technologically it is very difficult to manufacture good quality 1D
quantum wires with confinement in both spatial directions. They can be
obtained from a 2D quantum well, fabricated by thin-film growth, by
lateral structuring using lithographic methods. The accuracy of this
method is, however, limited to some ten nanometers and thus the
electronic properties of samples constructed in this way typically
have a strong inhomogeneous broadening. Fortunately it appears
possible to achieve quasi-1D particles even without a rigorous
confinement in any of the spatial directions. This has been realised
in so called V and T-shaped quantum wires. V-shaped quantum wires are
obtained by self-organised growth in pre-patterned materials such
as chemically etched V-shaped grooves in GaAs substrates. The T-shaped
quantum wire, first proposed by Chang et al.\ \cite{Chang}, forms at
the intersection of two quantum wells and is obtained by the cleaved
edge over-growth (CEO) method, a molecular-beam epitaxy (MBE)
technique. The accuracy of this method is extremely high and allows
fabrication of very thin (less than the Bohr radius of an exciton)
wires with small thickness fluctuations. These structures are
currently the subject of intensive research and have been realised by
several groups \cite{N2N4}-\cite{R}.

Experimentalists try to optimise the geometry and the materials in
order to increase the binding energy of the excitons, $E_b$, and the
confinement energy, $E_{con}$ for possible room temperature
applications. Up until now, the most popular material studied
experimentally has been GaAs/Al$_{x}$Ga$_{1-x}$As. Increasing the Al
molar fraction, $x$, should lead to bigger $E_b$ and $E_{con}$ but,
unfortunately, for larger $x$ the interfaces get rougher which
degrades the transport properties. Thus optimised geometries for lower
values of $x$ become more relevant.

The confinement energy, $E_{con}$, is the energy difference between
the lowest excitonic state in the wire and the lowest excitonic state
in the 2D quantum well. It can be directly measured as the difference
between the photoluminescence peaks obtained in a quantum wire (QWR)
and a quantum well (QW). It is, however, not possible to measure the
exciton binding energy directly. Its value has to be obtained from a
combination of experimental data and one-particle calculations of
electron and hole energies in a wire. There has been a disagreement
between the purely theoretical values \cite{Nonvar1}--\cite{Var3} and
those obtained from a combination of experimental data and theoretical
calculations.  The confinement energies, however, tend to agree
between experiment and purely theoretical calculations, suggesting
that experiment, using combined methods where errors tend to
accumulate, usually overestimates the binding energy.

\begin{figure}
      \begin{center}
        \leavevmode
      \epsfxsize=6.0cm
      \epsfbox{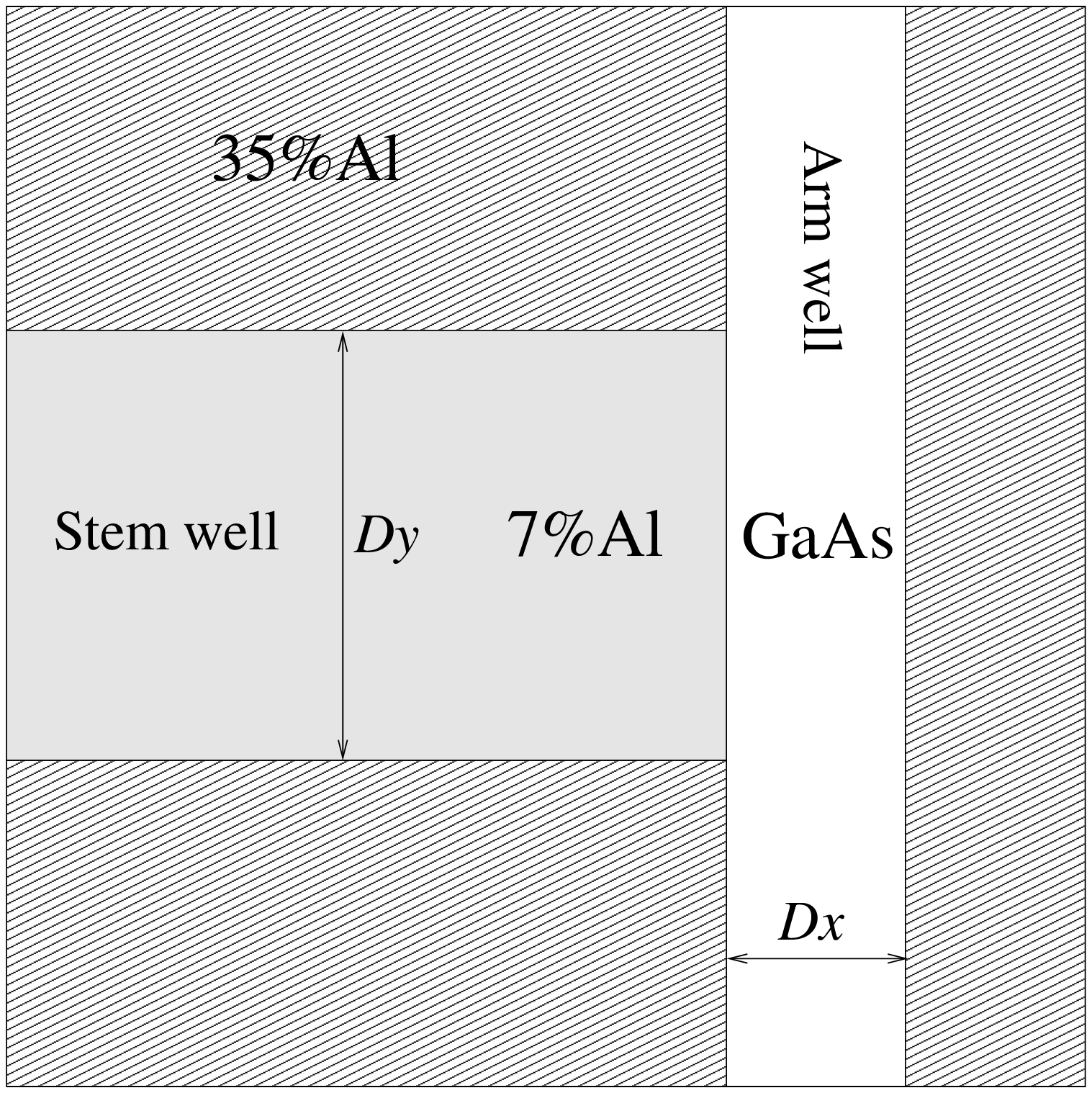}
        \end{center}
  \caption{Shape of the asymmetric T-shaped wire with notations. }
\label{asympot}
\end{figure}

For the 5-nm scale symmetric GaAs/AlAs, Someya et al. \cite{S1S2}
reported the largest confinement energy for excitons in symmetric
wires (Figure \ref{sympot}), $E_{con}$=38 meV and $E_b$=27$\pm$3
meV. The largest confinement energy of any structure was reported by
Gislason et al. \cite{G1,G2} for their optimised wires. Using
asymmetric wells with different widths and Al content as in
Figure \ref{asympot}, they obtained an exciton confinement energy of 54
meV. Recently there has also been the first experimental realisation
of T-shaped wires using In$_{y}$Ga$_{1-y}$As/Al$_{0.3}$Ga$_{0.7}$As
\cite{N2N4}. The highest confinement energy reported for this
structure is 34 meV, which is very close to the GaAs/AlAs result, and
the quality of the structure can be much higher than for the GaAs/AlAs
case.

Laser emission from the lowest exciton state in atomically smooth
semiconductor quantum wires was first observed by Wegscheider et
al. \cite{W} in symmetric, T-shaped quantum wires made on the
intersection of two 70 {\AA} GaAs quantum wells surrounded by AlGaAs
with the Al fraction $x=0.35$. Recently the same group obtained
excitonic lasing in a 60 {\AA}/140 {\AA} asymmetric quantum wire with
a 7\% Al filled Stem well (see Figure \ref{asympot} \cite{R}). They
reported an interesting observation of two-mode lasing in this
structure. Under strong excitation they achieved simultaneous lasing
from two levels in the quantum wire. There is a switching between
those two lasing modes as the temperature or pumping rate is
changed. A simple rate equation model \cite{rate} gives very good
agreement with experimental data, which suggests that we have lasing
from two different states in the quantum wire.

All calculations published to date which include the Coulomb
interaction between the electron and hole have only examined the
ground state exciton. They have used either variational methods
\cite{Var1}--\cite{Var3} or other approximations
\cite{Nonvar1,Nonvar2} and were performed only for symmetric wires and
for very limited cases realised experimentally in the early days of
T-shaped wire manufacturing. With the growing experimental realisation
of these structures as well as the interesting report of lasing
phenomena there is a need for accurate two-body calculations, treating
on an equal footing the single-particle potential and the Coulomb
interaction, of both the ground and excited states in the structure.

Excited states seem to be very important for the operation of
excitonic lasers \cite{R}. Calculations of energies, oscillator
strengths for radiative recombination (i.e, how the various states
couple to photons) as well as the full wave functions for the whole
spectra of interest would be very beneficial for understanding the
origins of certain transitions and effects. This could help in the
design of lasers with better properties and higher maximum
temperatures for excitonic lasing. The goal is to design excitonic
lasers which can operate at room temperature. Also, performing highly
accurate calculations of the ground state exciton in QWR and the
corresponding QW enables $E_b$ and $E_{con}$ to be obtained for
different geometries (both symmetric and asymmetric) for a wide range
of well widths and Al content, $x$. Such data are of great importance
for the optimisation of the structures.

Our method is based on an exact numerical solution of the
Schr\"{o}dinger equation in a certain basis within the effective mass
approximation.  The method is not restricted to a given number of
excited states and we can calculate as many of them as required. For
some structures we have calculated up to 100 excited states. We
perform calculations for a very wide range of T-shaped wires. In
section \ref{model} of the paper the numerical method is discussed in
detail while in section \ref{results} we present the results. There we
first study the spectra and wave functions and present a discussion of
the nature of the various excited states. Finally we discuss
$E_{con}$, $E_b$ and the difference between the ground-state exciton
energy and the first excited-state energy, $E_{2-1}$, as a function of
well width $Dx$ and Al molar fraction $x$ for the symmetric and asymmetric
quantum wires.

\section {The model}
\label{model}

\begin{figure}
      \begin{center}
        \leavevmode
      \epsfxsize=6.0cm
      \epsfbox{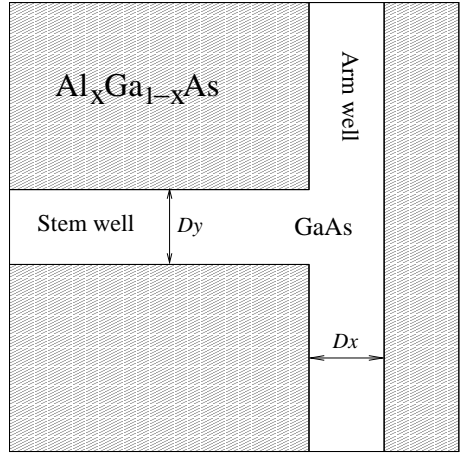}
        \end{center}
  \caption{Shape of the symmetric T-shaped wire with notations. }
\label{sympot}
\end{figure}

We use the effective mass approximation with an anisotropic hole mass
to describe an electron in a conduction band and a hole in a valence
band in the semiconductor structures under consideration. The
effective mass of the hole depends on the crystallographic direction
in the plane of the T-shaped structure. The electron and hole are in
the external potential of the quantum wire formed at the T-shaped
intersection of the GaAs/Al$_{x}$Ga$_{1-x}$As quantum wells. The so
called Arm quantum well is grown in the 110 crystal direction and
intersects with a Stem quantum well grown in the 001 direction (see
Figures \ref{sympot} and \ref{asympot}). In our model the crystal
directions 110, 001, and 110 correspond to $x$, $y$, and $z$
respectively. We consider symmetric quantum wires where the Arm and
Stem well are both of the same width, i.e, $Dx=Dy$, and are made of
GaAs. We also consider asymmetric wires where the Stem well is
significantly wider but filled with Al$_{x}$Ga$_{1-x}$As with a low Al
content to compensate for the reduction in confinement energy. Our
method is applicable to any structure regardless of its shape and
materials provided the external potential is independent of $z$

The value of the band-gap is different for the different materials
used in the well construction. This gives rise to the potential
barriers at the interfaces between the GaAs, Al$_{x}$Ga$_{1-x}$As and
InGaAs which take different values for electrons and holes. In our
model the electron and hole are placed in external potentials
$V_e(x,y)$ and $V_h(x,y)$, respectively, and interact via the Coulomb
interaction. We choose the potential in GaAs to be zero and calculate
all potentials in other materials with respect to this level. The
external potential is independent of $z$ in all cases. Sample
geometries considered in this work are shown in Figs. \ref{sympot} and
\ref{asympot}. Using the above model, after the separation of the
centre of mass and relative motion in the $z$ direction, the system is
described by the following Hamiltonian:
\begin{eqnarray}
 H =  
 -\frac{\hbar^2}{2m_e}\nabla^{2}_{x_{e},y_{e}}
 -\frac{\hbar^2}{2m_{hx}}\nabla^{2}_{x_{h}}
 -\frac{\hbar^2}{2m_{hy}}\nabla^{2}_{y_{h}}
 -\frac{\hbar^2}{2\mu_z}\nabla^{2}_{z}+
\nonumber\\
V_{e}(x_{e},y_{e}) + V_{h}(x_{h},y_{h}) 
 -\frac{e^2}{\sqrt{(x_{e}-x_{h})^{2}+(y_{e}-y_{h})^{2}+z^{2}}}, \nonumber
\end{eqnarray}
where $z=z_e-z_h$ and
$\frac{1}{\mu_z}=\frac{1}{m_e}+\frac{1}{m_{hz}}$. The wave function
associated with the centre of mass motion in the $z$ direction is a
plane wave and this coordinate can be omitted from the problem.

\subsection {Numerical method for calculating quantum wire exciton states}

We calculate the ground and excited states in the structures of
interest by a direct diagonalisation method. Due to the complexity of
the external potential with its limited symmetry and sharp edges, none
of the standard basis sets seem appropriate. We use the following
basis set:
\begin{eqnarray}
\label{basis-set}
\lefteqn{\psi(x_e,y_e,x_h,y_h,z_e-z_h)=} \nonumber\\
& & \sum_{i,j,k}c_{i,j,k}sin(z\frac{k\pi}{L_z}-
\frac{k\pi}{2})\chi^e_i(x_e,y_e)\chi^h_j(x_h,y_h), 
\end{eqnarray}
where $\chi^e_i(x_e,y_e)$/$\chi^h_j(x_h,y_h)$ are electron/hole
single-particle wave functions for a T-shaped potential without the
electron-hole Coulomb interaction. In the $z$ direction we introduce
hard wall boundary conditions and use a standing-wave basis set.

Our basis set does not obey the so called cusp condition \cite{Kato}
which is satisfied whenever two particles come together. The
divergence in the potential energy when the electron and hole come
together must be exactly cancelled by an opposite divergence in the
kinetic energy. The exact wave function must therefore have a cusp
when the electron and hole are coincident. Using a basis in which
every basis function obeys the cusp condition would reduce the size of
the basis set required. For an isotropic hole mass it would be very
easy to satisfy the cusp condition by multiplying the basis functions
by the factor $e^{-\Lambda\sqrt{(x_e-x_h)^2+(y_e-y_h)^2+z^2}}$ which
is just the hydrogenic wave function. Unfortunately there is no
analytical solution when we introduce the anisotropic hole mass. Thus
we choose not to satisfy the cusp condition and therefore have to use
a larger basis set.

The diagonalisation is performed using a NAG library routine.
Convergence is usually achieved with a basis set containing 20 of each
of the single-particle wave functions and 20 standing waves in the $z$
direction. Thus $20\times20\times20$ = 8000 basis functions are needed
which gives $20^6$ matrix elements. Only one quarter of the total
number needs to be calculated as interchanging $k_1$ and $k_2$ leaves
the matrix element unchanged while interchanging $i_1$ and $j_1$ with
$i_2$ and $j_2$ gives its complex conjugate. This still leaves a great
many matrix elements to be calculated. Thus to make the calculations
feasible the matrix elements need to be calculated very rapidly (See
Section \ref{matrix} .

\subsection{Computational method for calculating the single-particle
 wave functions}

\begin{figure}
      \begin{center}
        \leavevmode
      \epsfxsize=9.0cm
      \epsfbox{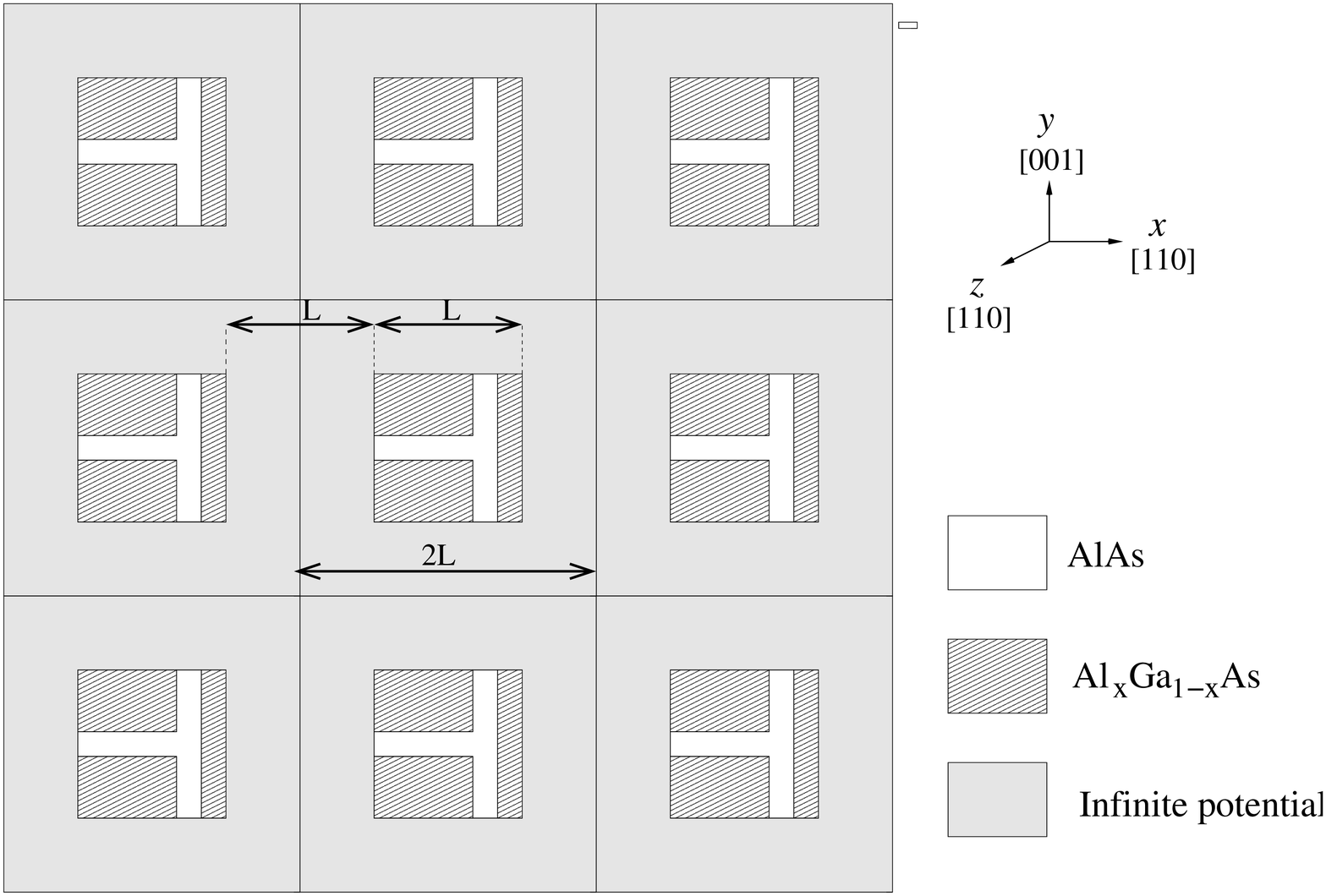}
        \end{center}
  \caption{Lattice used for calculations and notations. }
\label{lattice}
\end{figure}

The one-particle (electron and hole) wave functions,
$\chi^e_i(x_e,y_e)$ and $\chi^h_j(x_h,y_h)$ in a T-shaped external
potential are calculated using the conjugate-gradient minimisation
technique with pre-conditioning of the steepest descent vector. A
detailed explanation of this method can be found in reference
\cite{Payne}. We specify the external potential on a 2D grid and use
periodic boundary conditions in the $x$ and $y$ directions so that we
are able to use Fast Fourier Transform (FFT) methods to calculate the
kinetic energy in Fourier space while the potential energy matrix
elements are calculated in real space. The fast calculation of the
energy matrix elements is crucial as they have to be calculated many
times during the conjugate-gradient minimisation. The FFT provides
very fast switching between real and Fourier space and makes the
algorithm much more efficient, but the use of periodic boundary
conditions introduces the problem of inter-cell interactions in the
case of two particle calculations. To avoid this problem we place the
unit cell in the middle of another, larger unit cell of infinite
potential (see Figure \ref{lattice} and the Section \ref{matrix}).

We use plane waves as a basis set for the one-particle problem. Using
this method we can calculate as many as 50 states for the electron and
50 for the hole. Very good convergence with respect to the number of
plane waves and the size of the unit cell is obtained (see Section
\ref{accuracy}).

\subsection{Computational method for calculating the matrix 
elements}
\label{matrix}

The kinetic and potential energies are diagonal in this basis and are
obtained from the one-particle calculations. Thus only the Coulomb
matrix elements need to be calculated.

A Coulomb matrix element in the basis set (\ref{basis-set}) is a 5D
integral of the following form:
\begin{eqnarray}
\lefteqn{-\int\int\int\int\int dx_e dy_e dx_h dy_h dz  
sin(z\frac{k_2\pi}{L_z}- \frac{k_2\pi}{2})} \nonumber\\ 
& & \chi^{e*}_{i_2}(x_e,y_e)
    \chi^{h*}_{j_2}(x_h,y_h) 
    \frac{e^2}{\sqrt{(x_{e}-x_{h})^{2}+(y_{e}-y_{h})^{2}+z^{2}}}\nonumber\\ 
& & sin(z\frac{k_1\pi}{L_z}- \frac{k_1\pi}{2})\chi^e_{i_1}(x_e,y_e)
    \chi^h_{j_1}(x_h,y_h). \nonumber
\end{eqnarray}
This integral must be calculated numerically. Numerical integration
for so many dimensions is very slow and thus is not feasible for the
case of $20^6$ matrix elements. Thus another method has to be
introduced.

The above integral is of the form 
\begin{eqnarray}
\lefteqn{-\int\int\int\int\int dx_e dy_e dx_h dy_h dz} \nonumber\\
& & f_e(x_e,y_e)f_h(x_h,y_h)q(x_{e}-x_{h},y_{e}-y_{h},z)f_z(z). \nonumber
\end{eqnarray}
Where 
\begin{eqnarray}
\lefteqn{f_e(x_e,y_e)=\chi^{e*}_{i_2}(x_e,y_e)\chi^e_{i_1}(x_e,y_e)}, 
\nonumber\\
& & f_h(x_h,y_h)=\chi^{h*}_{i_2}(x_h,y_h)\chi^h_{i_1}(x_h,y_h), \nonumber\\
& &q(x_{e}-x_{h},y_{e}-y_{h},z)=
-\frac{e^2}{\sqrt{(x_{e}-x_{h})^{2}+(y_{e}-y_{h})^{2}+z^{2}}}.
\nonumber
\end{eqnarray}

Using the Fourier transform and the convolution theorem it can be
shown that the above integral is equal to:
\begin{eqnarray}
\label{lastform}
\int dz \sum_{G_x,G_y}F_e(-G_x,-G_y)*F_h(G_x,G_y)*Q(G_x,G_y,z).
\end{eqnarray}

Where $F_e$, $F_h$, $Q$ are the 2D Fourier transforms of the function
$f_e$ with respect to $x_e$ and $y_e$, $f_h$ with respect to $x_h$ and
$y_h$ and $q$ with respect to $x_e-x_h$ and $y_e-y_h$,
respectively. Thus the 5D integral can be reduced to a 1D integral
with respect to the $z$ variable and a 2D sum in Fourier space. The
$F_e$ and $F_h$ Fourier transforms can be easily calculated using FFTs
in real space after multiplication of the corresponding
$\chi^e_{i_1}(x_e,y_e)$ by $\chi^{e*}_{i_2}(x_e,y_e)$ for electrons
and $\chi^h_{i_1}(x_h,y_h)$ by $\chi^{h*}_{i_2}(x_h,y_h)$.
 
In order to use FFTs we need to introduce periodic boundary
conditions in the $x$ and $y$ directions as in the one-particle 
calculations. To eliminate interactions between particles in 
neighbouring cells, we place the unit cell in the middle of another,
bigger unit cell of infinite potential (see Figure \ref{lattice}).

The distance between the edges of successive small unit cells is
exactly the width of the small unit cell, $L$. We cut-off the Coulomb
interaction at a distance corresponding to the size of the small unit
cell. We therefor consider the following form of Coulomb interaction:
\begin{eqnarray}
\lefteqn{q(x_e-x_h,y_e-y_h,z)=} \nonumber\\
& & \left\{ \begin{array}{ll}
-\frac{e^2}{\sqrt{(x_{e}-x_{h})^{2}+(y_{e}-y_{h})^{2}+z^{2}}}
& \parbox{3cm}{if $x_e-x_h < L_x$ and $y_e-y_h < L_y$} \\
0 & \mbox{otherwise}.
\end{array}
\right. \nonumber
\end{eqnarray}
Particles interact only when their separations in the $x$ and $y$
directions are smaller than $L_x$ and $L_y$ respectively. The
separations of particles in neighbouring cells is always bigger than
the cut-off and thus they do not interact. Particles in the same unit
cell are always separated by less than that the cut-off distance due
to the infinite potential outside the small unit cell. Thus we take
into account all of the physical Coulomb interaction and completely
eliminate the interactions between images. In the numerical
implementation the infinite potential is replaced by a large but
finite potential. Thus the probability of the particle being outside
the small unit cell is effectively zero and we find that the results
do not depend on the value of this potential for values greater than
around three times the potential in the Al$_{x}$Ga$_{1-x}$As region.

The 2D Fourier transform of the 3D Coulomb interaction with a cut-off
cannot be done analytically. Thus we put the Coulomb interaction onto
a 2D grid as a function of relative coordinates $x_{e}-x_{h}$ and
$y_{e}-y_{h}$ for every $z$ value. The unit cell in relative
coordinates will go from $-L_x$ to $L_x$, and $-L_y$ to $L_y$
respectively. Then for every value of $z$ a 2D FFT is performed with
respect to $x_{e}-x_{h}$ and $y_{e}-y_{h}$ and the results stored in
the 3D array $Q(G_x,G_y,z)$. Since this is the same for every matrix
element the above calculation needs to be performed only once.

The calculations described by Eqn. \ref{lastform} need to be performed
for every matrix element. After $F_e(G_x,G_y)$ and $F_h(G_x,G_y)$ have
been calculated the summation over the reciprocal lattice vectors
$G_x$ and $G_y$ for every value of $z$ is performed. The remaining 1D
integral in the $z$ direction is done numerically, after interpolation
of data points, using a routine from the NAG library. The dependence
of the integrand on $z$ is found to be very smooth and thus not many
points are required to obtain accurate results.

\section {Results}
\label{results}

We perform the calculations for a series of T-shaped structures.  We
calculate energies, oscillator strengths and wave functions for the
first 20-100 two-particle states for symmetric and asymmetric wires.

For symmetric wires we consider the structure denoted by W which has
been experimentally studied by Wegscheider et al. \cite{W} and
consists of GaAs/Al$_{0.35}$Ga$_{0.65}$As 70 {\AA} quantum
wells. Then, keeping the rest of parameters constant, we vary the
quantum well width from 10 {\AA} to 80 {\AA} in steps of 10 {\AA} in
order to examine the width dependence of the various properties. We
also perform calculations for samples denoted by S1 and S2 studied by
Someya et al. \cite{S1S2} made of GaAs/Al$_{0.3}$Ga$_{0.7}$As (S1) and
GaAs/AlAs (S2) quantum wells of width around 50 {\AA}. For the
GaAs/AlAs case we again vary the well width from 10 {\AA} to 60
{\AA}. Then we take an intermediate value of the Al molar fraction,
$x=0.56$, and vary the well width from 10 {\AA} to 60 {\AA} in order
to examine the dependence on the well width as well as Al
content. Finally we perform calculations for 35 {\AA}-scale
In$_{0.17}$Ga$_{0.83}$As/Al$_{0.3}$Ga$_{0.7}$As (denoted by N4) as
well as for 40 {\AA}-scale
In$_{0.09}$Ga$_{0.91}$As/Al$_{0.3}$Ga$_{0.7}$As (denoted by N2)
samples as studied experimentally by Akiyama et al. \cite{N2N4}.

For asymmetric structures we consider the wire studied experimentally
by Rubio et al. \cite{R} which consists of a 60 {\AA}
GaAs/Al$_{0.35}$Ga$_{0.65}$As Arm quantum well and a 140 {\AA}
Al$_{0.07}$Ga$_{0.93}$As/Al$_{0.35}$Ga$_{0.65}$As Stem quantum well. We
vary the width of the Arm quantum well from 50 {\AA} to 100 {\AA}. We
also perform calculations for the asymmetric structure studied by a
different group \cite{G1,G2} which consists of a 25 {\AA}
GaAs/Al$_{0.3}$Ga$_{0.7}$As Arm quantum well and a 120 {\AA}
Al$_{0.14}$Ga$_{0.86}$As/Al$_{0.3}$Ga$_{0.7}$As Stem quantum well.
  
In the first part of this section we present the spectra for symmetric
and asymmetric quantum wires with the positions of 2D exciton, 1D
continuum (unbound electron and hole both in the wire) and 1De/2Dh
continuum (unbound electron in the wire and hole in the well) states
as well as pictures of representative wave functions.  This allows us
to discuss the nature of the excited states in the structures. In the
second part we discuss the trends in confinement and binding energy
and the separation in energy between the ground and the first excited
states as a function of the well width and Al fraction.

We use a static dielectric constant $\epsilon$=13.2 and a conduction
band offset ratio $Q_c=\Delta E_{cond}/\Delta E_g$ of 0.65. For the
difference in bandgaps on the GaAs/Al$_x$Ga$_{1-x}$As interface we use
the following formula: $\Delta E_g=1247\times x$ meV for $x<0.45$ and
$1247 \times x+1147 \times (x-0.45)^2$ meV for $x>0.45$. For the
electron mass we use $m_e = 0.067 m_0$ while for the hole mass $m_{hx}
= m_{hz} = m_{h[110]} =0.69-0.71 m_0$ and $m_{hy}=m_{h[001]}=0.38 m_0$
($m_0$ is the electron rest mass). For the
In$_{0.09}$Ga$_{0.91}$As/Al$_{0.3}$Ga$_{0.7}$As
(In$_{0.17}$Ga$_{0.83}$As/Al$_{0.3}$Ga$_{0.7}$As) we use parameters
from reference \cite{N2N4}: for the electron $m_e=0.0647(0.0626)m_0$,
for the hole $m_{hy}=m_{hh[001]}=0.367(0.358)m_0$ and
$m_{hx}=m_{hz}=m_{h[110]}=0.682(0.656)m_0$, $\Delta E_g$=464(557) meV
and the band offset was assumed to be 65\% in the conduction and 35\%
in the valence band.

\subsection{Excited states}
\subsubsection{Symmetric wires}
\label{symexstates}

\begin{figure}
      \begin{center}
        \leavevmode
      \epsfxsize=17.5cm
      \epsfbox{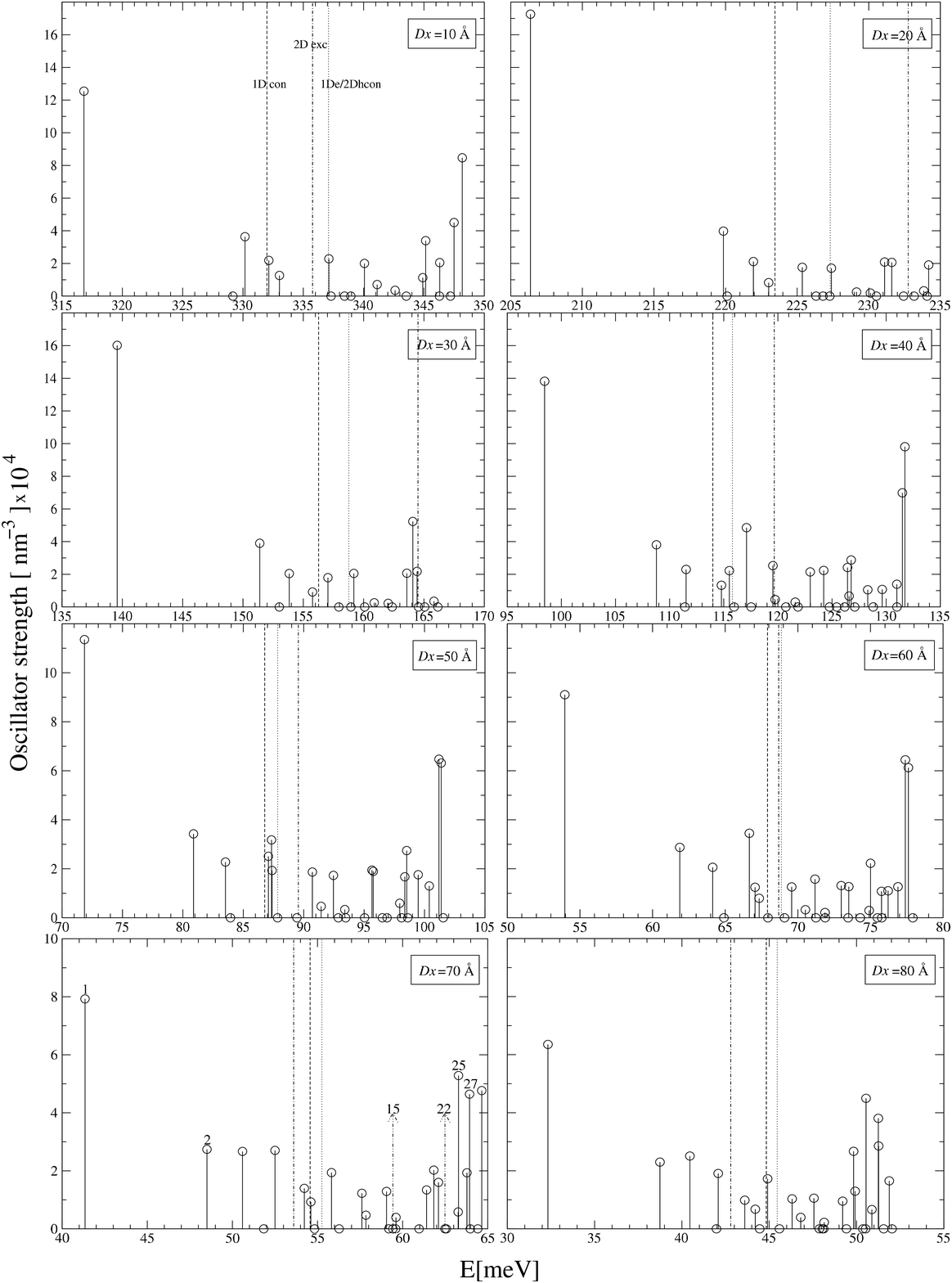}
        \end{center}
\caption{Oscillator strength versus energy for the lowest 20 states
in a symmetric T-shaped structure for different well widths $Dx$. }
\label{symspec}
\end{figure}

In Figure \ref{symspec} we show spectra (the oscillator strength
versus energy) for the first 20 (30 in the case of the 70 {\AA} wire)
states for the GaAs/Al$_{0.35}$Ga$_{0.65}$As structure for well widths
from 10 {\AA} to 80 {\AA}. A dashed line shows the energy of the 1D
continuum, a dotted line that of the 1D electron and 2D hole continuum,
while the dotted-dashed line shows that of the quantum-well 2D
exciton. Because our system is finite in the $z$ direction, we obtain
only a sampling of the continuum states; below the continuum edge the
states are discrete.

Note that for the experimentally studied 70 {\AA} structure, the 2D
exciton has a lower energy than the completely unbound electron and
hole in the wire. The situation clearly depends on the well width and
the crossing point is between 60 and 70 {\AA}. For well widths of 60
{\AA} or smaller, the 1D continuum (1Dcon) is lower in energy that the
2D exciton (2Dexc) with the difference being maximal for a width of
around 20 {\AA}. For widths of 70 {\AA} or bigger, the 1Dcon is higher
in energy than the 2Dexc with the difference growing for increasing
well width. This effect might be significant for pumping T-shaped-wire
lasers. Free electrons and holes are excited in the whole area of both
wells and thus, when the 2D exciton has a lower energy than the 1D
continuum, formation of the 2D excitons is energetically favourable.
These excitons can recombine in a well instead of going to the wire
and forming a 1D exciton. Clearly it is more efficient to have the 1D
continuum lower in energy than the 2D exciton.

By increasing the well width we obtain more states that are lower in
energy than the 1Dcon and 2Dexc beginning with two (ground and the
first excited) for the 10 {\AA} well, three for widths between 20-50
{\AA} and four states for larger widths.

\begin{figure}
      \begin{center}
        \leavevmode
      \epsfxsize=17.5cm
      \epsfbox{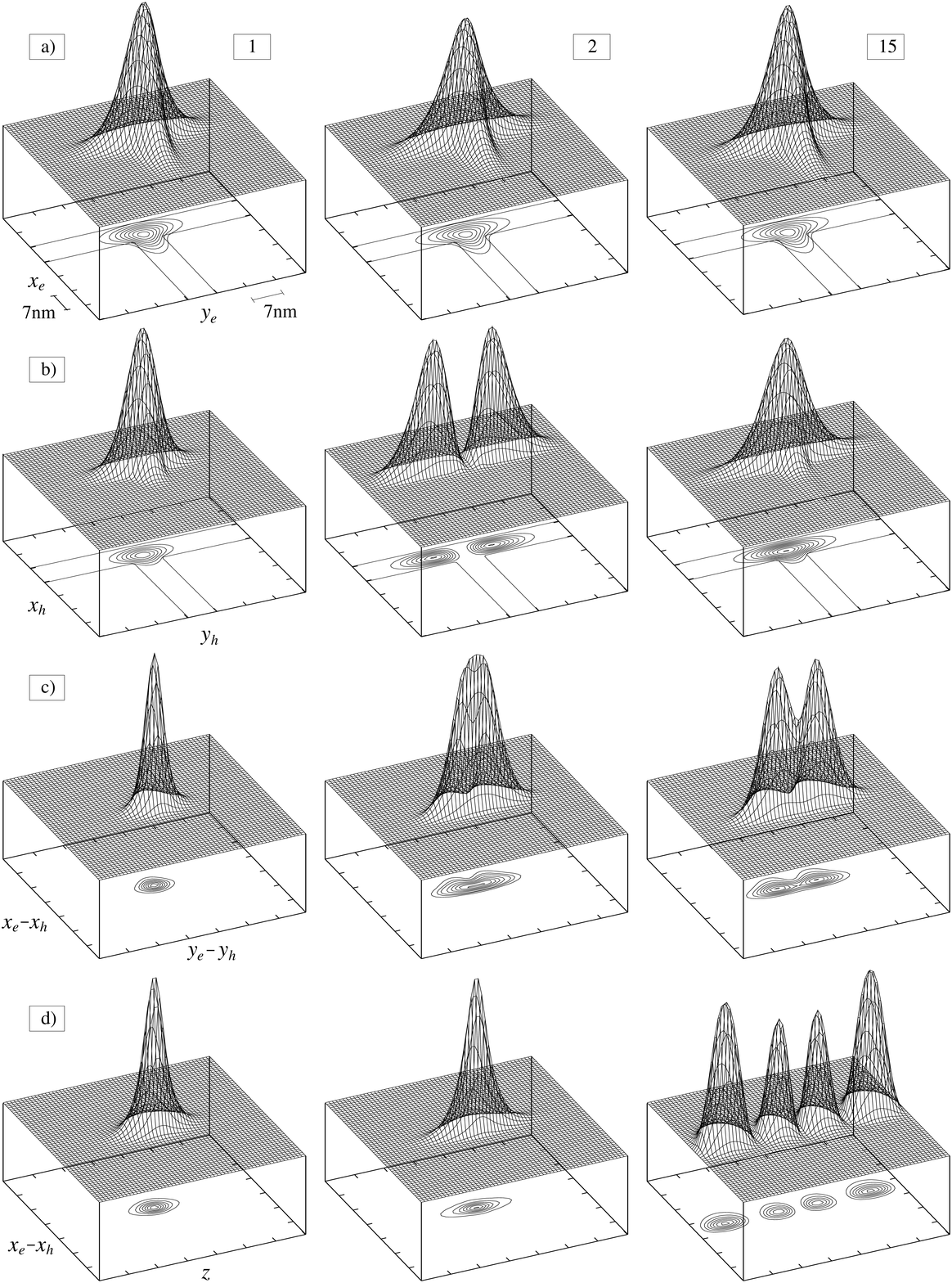}   
        \end{center}
\caption{Modulus squared of the two-particle wave function for the
ground (1), first excited (2) and the 15th (15) state in the symmetric
T-structure. Electron (a), hole (b) and the relative coordinates
$x_e-x_h$, $y_e-y_h$ (c), $x_e-x_h$, $z$ (d) probability densities are
shown. }
\label{sym70st1}
\end{figure}

\begin{figure}
      \begin{center}
        \leavevmode
      \epsfxsize=17.5cm
      \epsfbox{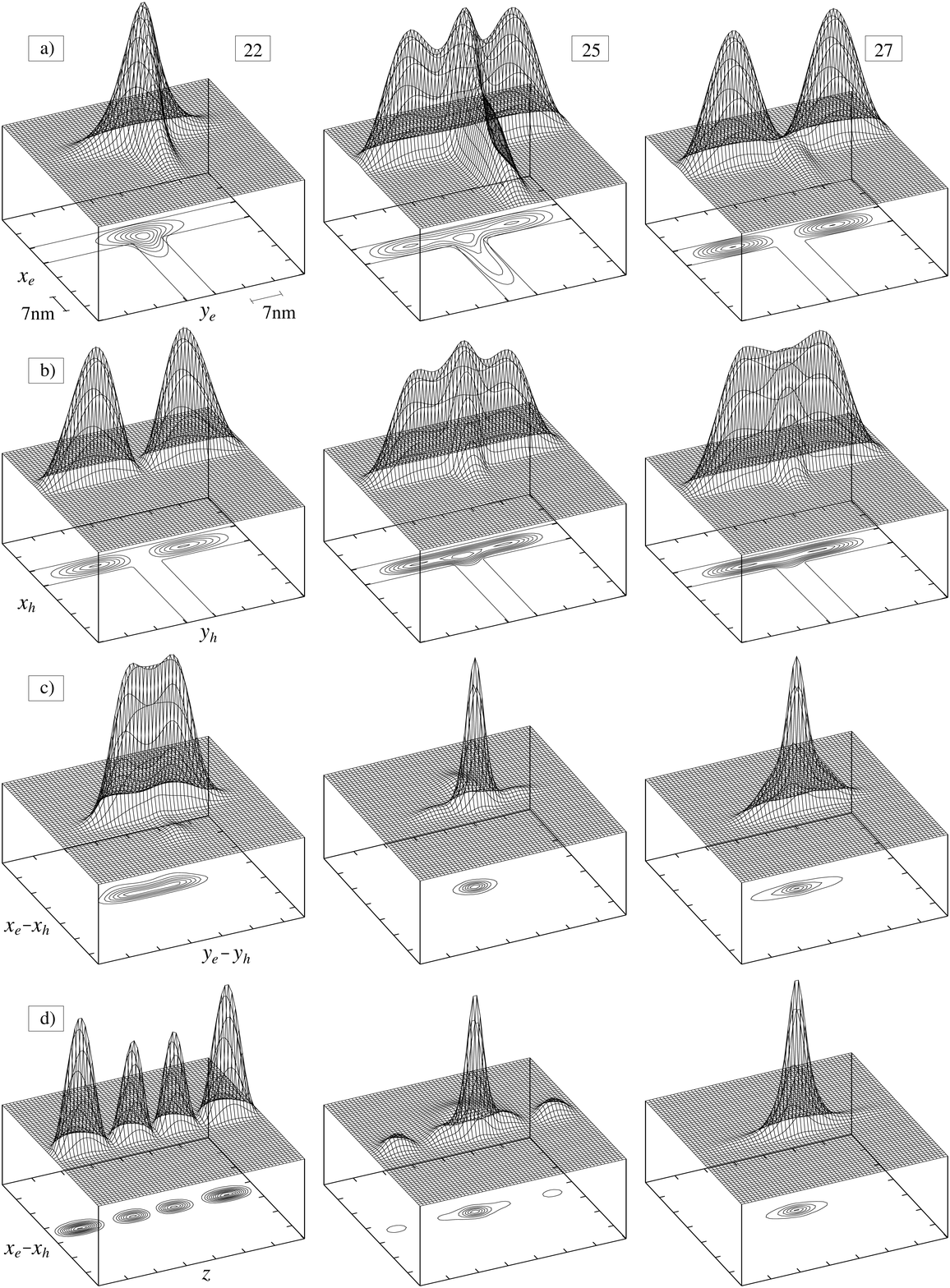}   
        \end{center}
\caption{Modulus squared of the two-particle wave function for the
22th, 25th and the 27th state in the symmetric T-structure. Electron
(a), hole (b) and the relative coordinates $x_e-x_h$, $y_e-y_h$ (c),
$x_e-x_h$, $z$ (d) probability densities are shown.}
\label{sym70st2}
\end{figure}

We now discuss the behaviour of $|\psi|^2$ for the 70 {\AA} case. The
wave functions depend on five spatial coordinates and thus various
cuts in 5D space are presented in Figures \ref{sym70st1} and
\ref{sym70st2}: a) the electron $x_e$, $y_e$ position after averaging
over the hole position, b) the hole $x_h$, $y_h$ position after
averaging over the electron position, and relative coordinates after
averaging over the centre of mass position c) the $x_e-x_h$, $y_e-y_h$
relative coordinates for $z_e-z_h=0$ and d) the $x_e-x_h$, $z$ relative
coordinates for $y_e-y_h=0$.

For the ground state we observe that the electron and hole are very
well localised in the wire with slightly more hole localisation. The
relative coordinate plots clearly show the bound exciton (Figure
\ref{sym70st1}(1)).

The electron in the first excited state is localised in the wire while
the hole already expands into the Arm well. The relative coordinate
pictures show that the electron and hole are bound and form an exciton
with an asymmetric shape. The size of the exciton is smallest in the
$x$ direction (the Stem well direction) and the exciton expands more
into the $y$ (the Arm well where the hole is expanded) and free $z$
directions (Figure \ref{sym70st1} (2)). The oscillator strength of
this state is about 1/3 of that of the ground state and the state
clearly takes the form of a 1D exciton with its centre of mass in the
T-wire.

It can be seen from the spectra (Figure \ref{symspec}) that there are
four states (apart from the ground state) with energies smaller than
1Dcon and 2Dexc. The nature of the 3rd and 5th states is very similar
to the 2nd one: the centre of mass is in the wire and the electron is
still well localised in the wire while the hole spreads into the wells
(into both the Arm and Stem wells for the 3rd state while only into
the Stem well for the 5th one). The relative coordinates show the
complex, asymmetric shape of this excitonic state and the oscillator
strength is again around 1/3 of the ground state exciton.

The 4th state with almost zero oscillator strength corresponds to a 1D
continuum. The electron and hole are both in the wire but the relative
coordinate pictures show an unbound exciton. Within the first 30
states we have 3 states of that nature: the 4th, 7th and 15th. The
15th state is shown in Figure \ref{sym70st1}: the electron and hole
are confined in the wire (a, b) and there are 3 nodes in the $z$
direction and 1 node in the $y$ direction. The other two states look
similar and differ only in the number of nodes. The energy of the 4th
state, which is the lowest 1Dcon state, turns out to be lower than the
real 1Dcon obtained from our one-particle calculations. This is due to
the finite size effects. Our method is very well converged with
respect to the cell size for the bound state and for the unbound ones
where at least one of the particles is in the well. However, for the
unbound continuum 1D states, the particles are very close in the $x,y$
plane because of the very small size of the wire and thus the
interaction is stronger. Consequently it does not decay as fast in
the $z$ direction as other states and thus we need a much bigger unit
cell in the $z$ direction to achieve convergence. There are however
only three such states within the 30 we examine and we know their true
energies from the preceding one particle calculations.

For further excited states up to the 25th, the electron, and thus the
centre of mass, is still localised in the wire while the hole is
taking up more and more energetic states in both wells, where energies
are quantised due to the finite size of the cell. Those states can be
divided into two groups depending on their relative coordinate nature:
excitonic-like states similar to the second state (Figure
\ref{sym70st1} (2)) and ionised states like the 22nd which is
represented in Figure \ref{sym70st2} (22). The oscillator strength of
the second group is zero (see Figure \ref{symspec}).

The 25th state (Figure \ref{sym70st2} (25)) is the first state where the
electron is de-localised in both wells, the relative coordinates
and the large oscillator strength shows that it is clearly an
excitonic-like state. It appears to be a 2D quantum well exciton state
scattered on the T shaped intersection. Its energy is thus higher than
that of a pure 2Dexc.

The 27th state is the 2D Arm-quantum-well-exciton state. It has higher
energy than the ground-state 2D exciton because the  electron and hole
wave functions occupy higher energy states than the ground state of the
well due to the presence of the T intersection.

The 30th state has a very similar nature to the 27th but the exciton
expands into the Stem instead of the Arm quantum well.

The 25th, 27th and 30th states all have large oscillator strengths
(around 3/4 of that of the ground state exciton). It is interesting to
note that between the ground state and those 2D
large-oscillator-strength states, there is a group of states with
relatively low oscillator strengths. The reason for this is that after
the ground state, there are states where either the wire-like electron
is bound to the well-like hole and thus they do not overlap enough to
give big contribution to the spectrum or they consist of a wire like
electron with an unbound hole.

Those quantum-well-like exciton states that scattered on the T-shaped
potential (like state 25) appear to be quite important for the
excitonic lasing because of their big oscillator strength. In \cite{R}
the authors reported two-mode lasing in an asymmetric wire where the laser
switches between the ground-state exciton and the other state whose
energy corresponds to the state from the tail of the above mentioned
states.

\subsubsection{Asymmetric wires}

\begin{figure}
      \begin{center}
        \leavevmode
      \epsfxsize=8.5cm
      \epsfbox{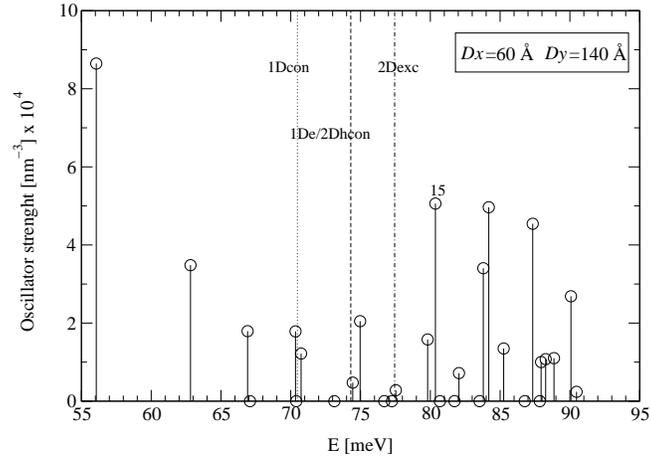}   
        \end{center}
\caption{Oscillator strength versus energy for the lowest 30 states in an
asymmetric T-shaped structure with $Dx$ = 60 {\AA}, $Dy$ = 140 {\AA}.}
\label{asymspec}
\end{figure}

The asymmetric wire that we study in detail consists of a 60 {\AA} or
56 {\AA} GaAs/Al$_{0.35}$Ga$_{0.65}$As Arm quantum well and a 140 {\AA}
Al$_{0.07}$Ga$_{0.93}$As/Al$_{0.35}$Ga$_{0.65}$As Stem quantum well
\cite{R}. The spectrum for the 60 {\AA} Arm case is shown in Figure
\ref{asymspec}. The nature of the states is very similar to the case
of the symmetric wire. The first two excited states are exciton-like
and have an electron confined in the wire while the hole spreads into
the well. All excited states up to the 20th have the electron confined
in the wire. The hole spreads to one or both quantum wells taking up
more energetic states in the well. The relative coordinates show
either an exciton-like wave function (states with nonzero oscillator
strength in the spectra of Figure \ref{asymspec}) or the case where a
hole is confined in the wire but is not bound to the electron
(states with zero oscillator strength in the spectra). Both groups
were discussed and shown for the symmetric wire.

The 21st and the 24th states (large oscillator strengths in the Figure
\ref{asymspec}) have an electron expanding into the Arm well. The
electron wave function has a node in the wire region. The hole
wave function spreads into the Arm well and has no node for the 21st
state and one node in the wire region for the 24th state. The relative
coordinates show the excitonic nature of these states.  Thus these
states correspond to those 2D excitonic states scattered on the wire.

\begin{figure}
      \begin{center}
        \leavevmode
      \epsfxsize=8.5cm
      \epsfbox{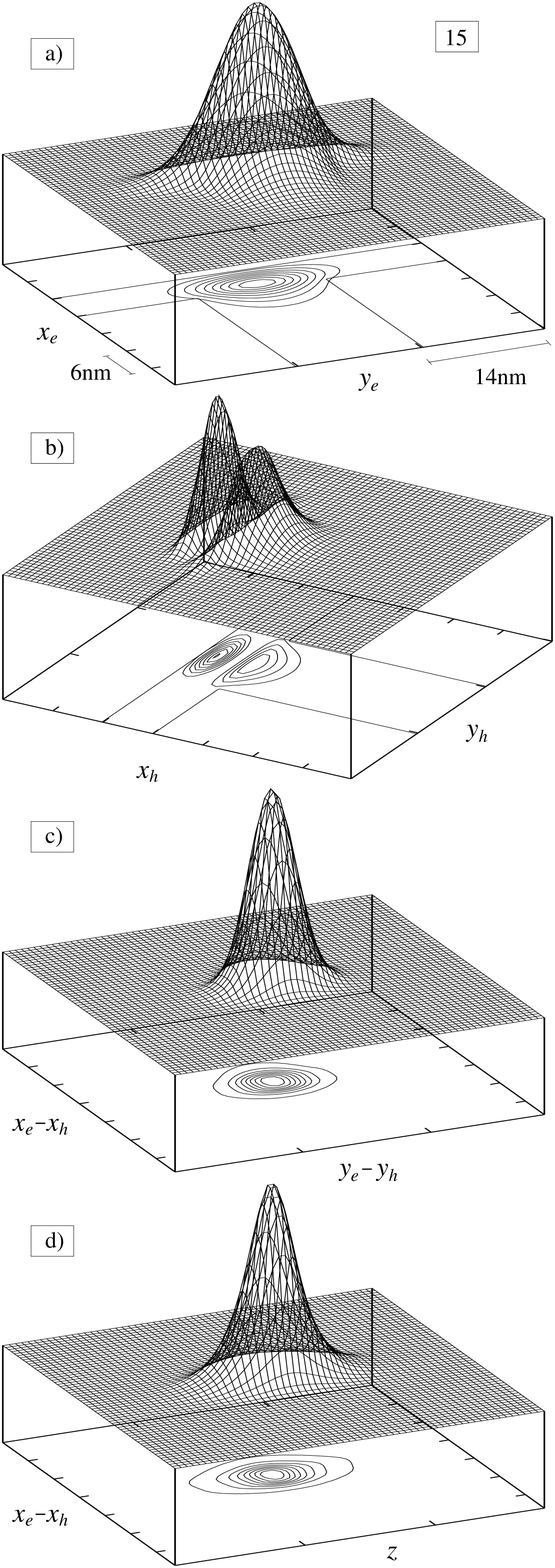}
        \end{center}
\caption{Electron (a), hole (b), and the relative coordinates
$x_e-x_h$, $y_e-y_h$ (c), $x_e-x_h$, $z$ (d) probability densities for
the 15th state in an asymmetric T-shaped structure with $Dx$ = 60
{\AA}, $Dy$ = 140 {\AA}.}
\label{asym15}
\end{figure}

For the asymmetric structure we observe one state (the 15th, see
Figure \ref{asym15}) which does not correspond to any state in the
symmetric case. The state is clearly excitonic-like with a large
oscillator strength and the relative coordinate plots show a very well
bound exciton. The electron is confined in the wire in the same way as
the ground state while the hole is clearly 1D-like, strongly confined
in the wire but in a different way. It has a node in the wire region.

\subsection{Trends in confinement and binding energies}

\subsubsection{Symmetric wires}

We calculate the exciton binding energy, $E_b=E_e+E_h-E_{1Dexc}$,
where $E_e$ and $E_h$ are the one-particle energies of an electron and
a hole, respectively, in the wire. We also calculate, using the same
method, the exciton energy in the quantum well, $E_{2Dexc}$, to obtain
the confinement energy of the 1D exciton, $E_{con}=E_{2Dexc}-E_{1Dexc}$,
in the wire.

\begin{figure}
      \begin{center}  
        \leavevmode
      \epsfxsize=8.5cm
      \epsfbox{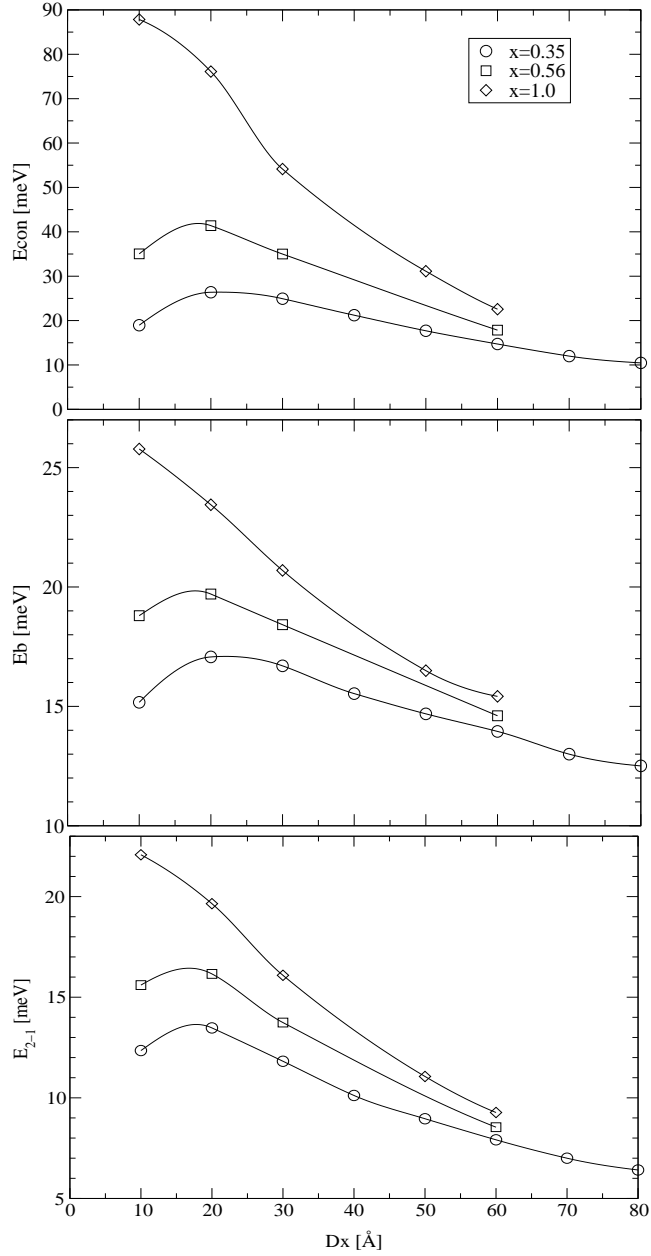}
        \end{center}
\caption{Confinement energy $E_{con}=E_{2Dexc}-E_{1Dexc}$, binding
energy of the ground-state exciton $E_b$, and the energy difference
between the ground state and the first excited state $E_{2-1}$ as a
function of the well width $Dx$ in a symmetric T-structure for three
different aluminium molar fractions $x$. }
\label{symEcEb}

\end{figure}

We perform calculations for a wide range of structural parameters. For
the GaAs/Al$_{x}$Ga$_{1-x}$As quantum wire we change the well width
from 10 {\AA} to 80 {\AA} for three different values of the Al content
$x$. The results are shown in Figure \ref{symEcEb}. It can be noticed
that for a well width bigger than 50 {\AA}, changing the Al content has
very little effect on the confinement and binding energies. The
difference in binding energy between the 60 {\AA}
GaAs/Al$_{0.35}$Ga$_{0.65}$As and the pure AlAs is only 1.5 meV. Thus
it seems more promising to change the well width rather than the Al
content for relatively wide wires. However, for thinner wires in the
range of 10-50 {\AA}, changing the Al content is much more profitable
then changing the well width. The difference in binding energies for
20 {\AA} wires with Al molar fractions of $x$=0.3 and $x$=1.0 is 6.4
meV. This increases to 10.6 meV when the width is reduced to 10 {\AA}.

$E_b$ and $E_{con}$ for Al contents of $x$=0.35 and $x$=0.56 both
approach a maximum for a well width between 10 {\AA} and 20 {\AA}. The
maximum values for $x$=0.35 are $E_{bmax}=17.1$ meV, $Econ_{max}=26.4$
meV and for $x$=0.56 they are $E_{bmax}=19.7$ meV, $Econ_{max}=41.4$
meV. For the $x$=1.0 case, the curve does not have a maximum in the
region for which calculations has been performed but we consider going
to wells thinner than 10 {\AA} as practically uninteresting. Thus the
maximum energies are for $Dx$=10 {\AA} and they are $E_{bmax}=25.8$ meV and
$Econ_{max}=87.8$ meV. 

$E_{con}$ increases much more rapidly than $E_b$ when the well width
is progressively reduced. The curves cross for a well width between 60
{\AA} and 70 {\AA}, i.e. for widths of 60 {\AA} or smaller, $E_{con}$
is greater that $E_b$ which means that the 1D continuum is lower in
energy than the 2D exciton (as we discussed in section
\ref{symexstates}) with the difference having a maximum at around 20
{\AA}. For widths of 70 {\AA} or bigger, $E_b$ is greater than
$E_{con}$ with the difference growing for increasing well width. We
also consider the difference in energy between the ground state
exciton in the wire and the first excited state as a function of the
well widths. For the experimentally realised $Dx$ = 70 {\AA} case,
this difference is $E_{2-1}=7.0$ meV and the maximum value for $Dx$ =
10 {\AA} is $E_{2-1max}=13.5$ meV. The maximum value for the GaAs/AlAs
at $Dx$ = 20 {\AA} is 22 meV.

Although pure AlAs gives the biggest potential offsets and thus the
biggest binding and confinement energies, the GaAs/AlAs interfaces are
not very smooth, which influences the transport properties. Thus new
materials have to be proposed. Two structures based on InGaAs have
been manufactured and measured \cite{N2N4}: 35 {\AA}-scale
In$_{0.17}$Ga$_{0.83}$As/Al$_{0.3}$Ga$_{0.7}$As (N4) and 40 {\AA}-scale
In$_{0.09}$Ga$_{0.91}$As/Al$_{0.3}$Ga$_{0.7}$As (N2). The results of
calculations for these structures are presented in Table \ref{table}.
It can be seen that energies for the sample N4 are almost exactly
the same as for the GaAs/AlAs sample S2 suggesting that these
materials might be very good candidates for structures with large exciton
confinement and binding energies.

\subsubsection{Asymmetric wires}

In order to increase binding and confinement energies, the asymmetric
T-shaped structure was proposed and realised by two groups
\cite{R,G1,G2}. 

\begin{figure}
      \begin{center}  
        \leavevmode 
      \epsfxsize=8.5cm 
      \epsfbox{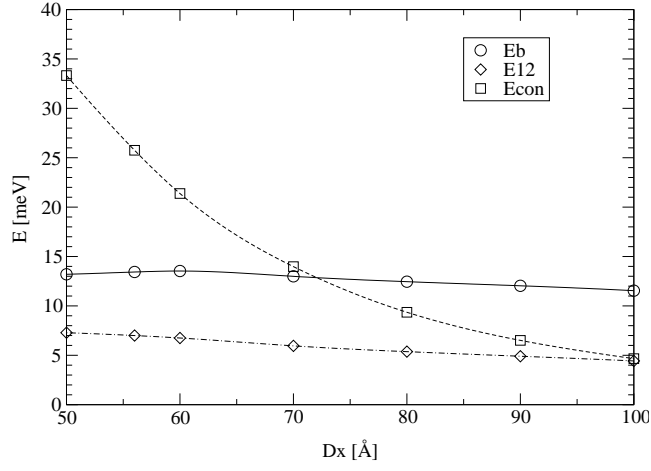} 
        \end{center} 
\caption{Confinement energy $E_{con}=E_{2Dexc}-E_{1Dexc}$, binding
energy of the ground state exciton $E_b$, and the difference between the
ground and the first excited state $E_{2-1}$ for an asymmetric
wire as a function of the well width $Dx$, where $Dy$ = 140 {\AA}.}
\label{nasEbEc}
\end{figure}

We calculate $E_b$ and $E_{con}$ for the 60 {\AA}/140 {\AA} structure
with the Stem quantum well filled with 7\% Al in order to compare with
experiment \cite{R} and then we vary the width of the Arm well from 50
to 100 {\AA}. One can see from Figure \ref{nasEbEc} that the binding
energy is almost independent of the Arm well width in this region,
changing only from the maximum value of 13.5 meV for $Dx$ = 60 {\AA}
to 11.5 meV for $Dx$ = 100 {\AA}. The binding energy for the 60 {\AA}
symmetric wire with the same $x$=0.35 Al mole fraction is 13.9 meV - a
bit bigger than for the asymmetric structure. In contrast, the
confinement energy, $E_{con}$, changes rapidly with the width of the
Arm well from 4.7 meV for $Dx$ = 100 {\AA} up to 33.3 meV for $Dx$ =
50 {\AA}. For Arm-well widths of 60 {\AA} or bigger, the 2D
quantum-well exciton in the Arm well has a lower energy than that for
the Stem well, thus the confinement energy is calculated with respect
to the Arm well exciton. For the 50 {\AA}-wide Arm well, the 2D
exciton has higher energy than for the Stem quantum well and thus the
confinement energy is calculated with respect to the Stem quantum
well. Therefore 33.3 meV is the highest confinement energy for this
Stem well and changing the Arm well would have no effect. Thus the 60
{\AA}/140 {\AA} structure is well optimised and its confinement
energy, $E_{con}$, is 21.4 meV which is much bigger than that of 14.7
meV for the 60 {\AA} symmetric wire.

The highest confinement energy so far reported is for an asymmetric
GaAs/Al$_{0.35}$Ga$_{0.65}$As wire with a 25 {\AA} Arm quantum well and
a 120 {\AA} Stem quantum well filled with 14\% Al \cite{G1,G2}. The
experimentally obtained $E_{con}$ for this structure is 54 meV.  Our
calculations however give only 36.4 meV which is still the highest
among experimentally obtained structures but much lower than 
reported by the authors. Our calculation of $E_{con}$ for five different
experimentally realised structures agree very well with the
experimental values and thus it is very probable that the value of 54 
meV is overestimated. The binding energy from our calculations is only
14.6 meV for this structure.

We can conclude from our results that the optimised asymmetric
structure does not lead to a bigger exciton binding energy than the
symmetric ones with the same parameters. The confinement energy is
considerably enhanced and this effect, which can be measured directly,
has often been used to infer that the binding energy is increased.
However, our results show that no such relationship holds between the
confinement and binding energies.  Thus the biggest confinement energy
of any structure constructed so far of 36.4 meV does not lead to the
biggest binding energy. Indeed, the binding energy of 14.6 meV is
smaller than the 16.5 meV reported for the GaAs/AlAs 50 {\AA}-scale
symmetric structure \cite{S1S2} where the confinement energy should be
only 31.1 meV. It is also smaller than expected for a symmetric 25
{\AA}-scale structure with the same parameters (16.0-16.5 meV). Thus
asymmetric structures could be useful for applications where a large
confinement energy is required but appear to be less suitable than
symmetric wires for applications where large binding energies are of
interest.

\subsubsection{Comparison with experiment and other calculations.}

The comparison between experiment and other published calculations is
presented in Table \ref{table}. The confinement energy of the exciton
can be directly measured experimentally. Although, due to the strong
inhomogeneous broadening of the photoluminescence peaks, the accuracy
of this number is not very high, it is the only experimentally proven
quantity we can refer to. The experimental binding energy needs to be
calculated using both experimental data and one-particle calculations
and thus errors might accumulate. Other theoretical methods which we
refer to obtain the ground state exciton energy using variational
techniques \cite{Var1}-\cite{Var3} (they differ in the form used for
the variational wave functions). There are also two non-variational
calculations for the ground state exciton \cite{Nonvar1,Nonvar2}.

Our results for the confinement energy of the ground state exciton
$E_{con}$ agree very well with experimental values for samples S1, N2
and N4 to an accuracy of 1\%, 6\% and 8\% respectively. This is indeed
very good agreement taking into account the strong inhomogeneous
broadening of the peaks they present. The spectral linewidth of the
photoluminescence peaks according to the authors is around 15 meV which
corresponds to a thickness fluctuation of about 3 {\AA} for N2 and N4
\cite{N2N4}. For the S1 and S2 samples the authors estimate the
experimental error due to the inhomogeneous broadening as 2 meV.
Agreement between our calculations and experiment is not as good for
the S2 sample but for this case additional effects are present.
For example, AlAs barriers give much less smooth interfaces than the
lower Al fraction samples and this is not taken into account in our
model. There is also very good agreement (better then 7\%) between our
results and the experimental measurement \cite{R} for asymmetric wire
R. The earlier $E_{con}$ published by this group for the symmetric
structure W is probably slightly overestimated.

There are only two calculations published for the confinement energy.
They are based on variational methods and were performed only for
sample W. Variational method 2 \cite{Var2} uses a wave function which
takes into account correlation in all spatial direction and the
agreement with our results is very good for the confinement energy but
not so good for the binding energy.

The variational method proposed by Kiselev et al. \cite{Var3} and
denoted here by ``3'' has a trial wave function which has only $z$
dependence in the correlation factor. Their binding energy for the
sample W differs by only 1 meV from our result but their value for the
confinement energy differs from ours. They
perform calculations of the binding energy for the whole range of
well widths, $Dx$, from 10-70 {\AA}. This can be compared with our results
in Figure \ref{symEcEb}. Their calculations, like ours, give the
maximum for $E_b$ and $E_{con}$ for a well width of around
20 {\AA}. Their binding energy is a bit bigger than the one
from our calculations. They obtained a maximum of $E_{b}=18.6$ meV which
is 1.5 meV higher than our result. However, their confinement energy
$Econ_{max}=33.0$ meV differs by 7 meV from our result. Their values of
$E_{con}$ are probably overestimated. They use the variational
technique to calculate the quantum wire exciton energy but the quantum
well exciton energy is taken from some other calculations of excitons
in quantum wells performed using a different method and with different
parameters, thus errors may accumulate.

The variational method 1 \cite{Var1}, which uses yet another form of
trial wave function, has been applied to samples S1 and S2 to calculate the
binding energy $E_b$. It agree quite well with our and other accurate
methods.

The binding energy we obtain shows excellent agreement with another
non-variational calculations by Glutsch et al.\cite{Nonvar1} (see
Table \ref{table}). They calculated the binding energy only for
samples W, S1 and S2 and thus unfortunately the confinement energy
cannot be compared. The method presented in reference \cite{Nonvar2}
gives much lower values for the binding energy than all other methods.

Despite some small differences, all of the theoretical methods give
much smaller values for $E_b$ than the experimental estimates. One has
to bear in mind, however, that the ``experimental'' values for $E_b$
(quoted in the Table \ref{table}) are in fact derived from a
combination of experimental data and associated theoretical modelling,
with inherent uncertainties.  Our results come from direct
diagonalisation and are very well converged. Therefore we believe that
the experimental binding energies are, in some cases, considerably
overestimated. The real binding energy is thus smaller than has been
claimed and the biggest value for any of the structures manufactured
so far is 16.5 meV for samples S2 and N4.

\subsubsection{Accuracy of the results}
\label{accuracy}
In our method the one-particle energies and wave functions are
calculated first. The one-particle energies are very well converged
with respect to all the variables such as unit cell size, number of
points on the grid and the number of plane waves to an accuracy of 0.1
meV. We use on average as many as 160 000 plane waves which
corresponds to $400 \times 400$ points on the grid ($200 \times 200$
in the small unit cell). We obtain excellent agreement between our
energies for the single electron and hole and those obtained by
Glutsch et al.\cite{Nonvar1}. For the 70 {\AA}, $x$=0.35 symmetric
quantum wire we obtain $E_e=47.09$ meV and $E_h=7.47$ meV while their
results are $E_e=47.2$ meV and $E_h=7.5$ meV. According to our
calculations there is only one electron state confined in the wire and
its confinement energy $E_{2D-1D}$ (i.e., the difference between
well-like and wire-like electron states) is 9 meV. This is in very
good agreement with other methods. L. Pfeifer et al. \cite{calc} using
eight band $\vec{k} \cdot \vec{p}$ calculations obtained a confinement
energy of 8.5 meV for the same structure. Kiselev et al. \cite{Var3}
using the so-called free-relaxation method obtained approximately the
same value of 9 meV.

These one-particle wave functions are then used as a basis set for the
two-particle calculations. The $E_{1Dexc}$ is very well converged with
respect to the number of points on the grid (as for the one-particle
calculations), and with the size of the basis set. Convergence is
usually achieved with about $20 \times 20 \times 20$ (8000) basis
functions. In order to minimise finite size effects we use quite big
unit cells (from 43 times the well width, $Dx$ for very thin wires
(10 {\AA}) to 7 times $Dx$ for the 80 {\AA} wire). The exciton energy
$E_{1Dexc}$ is converged to within about 0.2 meV and $E_{2Dexc}$ to
within 0.3 meV which gives an accuracy for $E_b$ of about 0.3 meV and
for $E_{con}$ of about 0.5 meV.

The other problem which can influence the accuracy of the results is
the uncertainty associated with the input parameters. The electron and
hole masses as well as the dielectric constant are standard but the
potential barriers vary a lot depending on the publication. We have
found quite different values of the potential offsets for the same
material interfaces in the literature. We have examined the influence of
this uncertainty on the final results by performing calculations for
the extrema of the sets of parameters found. The binding energy is
practically insensitive to those differences while the confinement
energy can differ by approximately 2 meV.

For the parameters that we are using, the results are converged to
within 0.3 meV for the binding and 0.5 meV for the confinement
energies. However, one needs to remember that these parameters are not
well calibrated and this could lead to an additional error in the
confinement energy of about 2 meV.

\section{Summary}

We have performed an exact diagonalisation within a finite basis set
of the Hamiltonian which describes an interacting electron-hole pair
in a T-shaped quantum wire. We have obtained the ground and excited
state energies and wave functions for this system. The first group of
excited states shows an $s$-like excitonic character where the
electron is localised in the wire but is bound to the hole which
spreads into one of the wells. Due to the fact that the electron and
hole are not localised in the same region, we have a group of low
oscillator-strength states just above the ground state. This group is
followed by a number of states with large oscillator strength which
are 2D excitonic states scattered on the T-shaped intersection. The
excitonic lasing from one of those states has been experimentally
observed \cite{R}. We have also performed a detailed study of the
exciton binding and confinement energies as a function of the well
width and Al molar fraction for symmetric and asymmetric wires. The
highest binding energy in any structure so far constructed is
calculated to be 16.5 meV which is much smaller than previously
thought. Our results have shown that for optimised asymmetric wires,
the confinement energy is enhanced but the binding energy is slightly
lower with respect to those in symmetric wires. For
GaAs/Al$_{x}$Ga$_{1-x}$As wires we have obtained an upper limit for
the binding energy of around 25 meV in a 10 {\AA} wide GaAs/AlAs
structure which suggests that other materials need to be explored in
order to achieve room temperature applications.
In$_{y}$Ga$_{1-y}$As/Al$_{x}$Ga$_{1-x}$As might be a good candidate.

\acknowledgments

We are pleased to thank P. Haynes, G. Rajagopal, A. Porter, Y. Mao and
G. McMullan for very beneficial discussions concerning the
computational techniques and L. N. Pfeiffer, A. Pinczuk, J. Rubio for
stimulating discussions and communicating experimental results prior
to publication. M.H.S acknowledge financial support from Lucent
Technologies, Trinity College Cambridge and an ORS award.

\end{multicols}


\begin{table}
\caption{Binding energy, $E_b$ and the confinement
energy $E_{con}=E_{1Dexc}-E_{2Dexc}$ in meV of the QWR exciton for five
different samples $W, S_1, S_2, N_2, N_4$ obtained from different methods.}
\begin{tabular}{lcccccccccccccc}
&\multicolumn{2}{c}{W\tablenote{Sample and experimental values from
Ref.\cite{W}.}}
&\multicolumn{2}{c}{$S_1$\tablenote{Sample and experimental values from
Ref.\cite{S1S2}.}}
&\multicolumn{2}{c}{$S_2$\tablenotemark[2]}
&\multicolumn{2}{c}{$N_2$\tablenote{Sample and experimental values from
Ref.\cite{N2N4}.}}
&\multicolumn{2}{c}{$N_4$\tablenotemark[3]}
&\multicolumn{2}{c}{$R$\tablenote{Sample and experimental values from
Ref.\cite{R}.}}
&\multicolumn{2}{c}{$G$\tablenote{Sample and experimental values from
Ref.\cite{G1,G2}.}}\\
Method&$E_b$&$E_{con}$&$E_b$&$E_{con}$&$E_b$&$E_{con}$&$E_b$&$E_{con}$&$E_b$
&$E_{con}$&$E_b$&$E_{con}$&$E_b$&$E_{con}$\\
\tableline
 Exp\tablenote{$E_{con}$ is obtained experimentally from the shift between
QW and QWR exciton lines. The $E_b$ is obtained indirectly from
experimental measurement of the QWR exciton line and one-particle
calculations. } &17&17&17&18&27&38&&28&&34&13.8&23&&54 \\
 This work&13&12&14.3&17.8&16.5&31.1&12.1&26.3&16.5&31.2&13.5&21.4&14.6&36.4\\
 Nonvar1\tablenote{Results of calculations from
Ref.\cite{Nonvar1}.}&13.2&&14.3&&16.4\\
 Nonvar2\tablenote{Results of calculations from
Ref.\cite{Nonvar2}.}&&&11.63&&13.9\\
 Var1\tablenote{Results of variational calculations from
Ref.\cite{Var1}.}&&&15&&18\\
 Var2\tablenote{Results of variational calculations from
Ref.\cite{Var2}.}&9.6&11.9\\
 Var3\tablenote{Results of variational calculations from
Ref.\cite{Var3}.}&12&14\\
 \end{tabular}
 \label{table}
\end{table}

\end{document}